\documentclass[iop,twocolappendix,appendixfloats,numberappendix]{emulateapj}
\usepackage{graphicx}
\usepackage[bookmarks=true,breaklinks=true]{hyperref}

\begin{document} 

\title{Accretion-powered pulsations in an apparently quiescent neutron star binary}
\shorttitle{Pulsations in a Quiescent LMXB}
\shortauthors{Archibald et al.}
\author{Anne M. Archibald\altaffilmark{1,*}}
\author{Slavko Bogdanov\altaffilmark{2}}
\author{Alessandro Patruno\altaffilmark{3,1}}
\author{Jason W. T. Hessels\altaffilmark{1,4}}
\author{Adam T. Deller\altaffilmark{1}}
\author{Cees Bassa\altaffilmark{1}}
\author{Gemma H. Janssen\altaffilmark{1}}
\author{Vicky M. Kaspi\altaffilmark{5}}
\author{Andrew G. Lyne\altaffilmark{6}}
\author{Ben W. Stappers\altaffilmark{6}}
\author{Shriharsh P. Tendulkar\altaffilmark{7}}
\author{Caroline R. D'Angelo\altaffilmark{3}}
\author{Rudy Wijnands\altaffilmark{4}}

\altaffiltext{1}{ASTRON, the Netherlands Institute for Radio Astronomy, Postbus 2, 7990 AA, Dwingeloo, The Netherlands}
\altaffiltext{2}{Columbia Astrophysics Laboratory, Columbia University, 550 West 120th Street, New York, NY 10027, USA}
\altaffiltext{3}{Leiden Observatory, Leiden University, PO Box 9513, NL-2300 RA Leiden, The Netherlands}
\altaffiltext{4}{Anton Pannekoek Institute for Astronomy, University of Amsterdam, Science Park 904, 1098 XH Amsterdam, The Netherlands}
\altaffiltext{5}{McGill University, 3600 University Street, Montreal, QC H3A 2T8, Canada}
\altaffiltext{6}{Jodrell Bank Centre for Astrophysics, School of Physics and Astronomy, The University of Manchester, Manchester M13 9PL, UK}
\altaffiltext{7}{Space Radiation Laboratory, California Institute of Technology, 1200 East California Boulevard, MC 249-17, Pasadena, CA 91125, USA}
\altaffiltext{*}{To whom correspondence should be addressed; E-mail: \protect \email{archibald@astron.nl}.}

\begin{abstract}
Accreting millisecond X-ray pulsars are an important subset of low-mass X-ray binaries in which coherent X-ray pulsations can be observed during occasional, bright outbursts (X-ray luminosity $L_X\sim 10^{36}\,\textrm{erg}\,\textrm{s}^{-1}$).  These pulsations show that matter is being channeled onto the neutron star's magnetic poles.
However, such sources spend most of their time in a low-luminosity, quiescent state ($L_X\lesssim 10^{34}\,\textrm{erg}\,\textrm{s}^{-1}$), where the nature of the accretion flow onto the neutron star (if any) is not well understood.
Here we report that the millisecond pulsar/low-mass X-ray binary transition object PSR~J1023+0038 intermittently shows coherent X-ray pulsations at luminosities nearly 100 times fainter than observed in any other accreting millisecond X-ray pulsar. 
We conclude that in spite of its low luminosity PSR~J1023+0038 experiences episodes of channeled accretion, a discovery that challenges existing models for accretion onto magnetized neutron stars.
\end{abstract}


\maketitle 

\section{Introduction}
Accretion is an important and ubiquitous process in astrophysics, but the nature of accretion flows is often puzzling.  Low-mass X-ray binaries (LMXBs) are transient objects in which material from a low-mass stellar companion accretes onto a compact object, either a black hole or a neutron star. At high accretion rates ($L_X\gtrsim 10^{35}\,\textrm{erg}\,\textrm{s}^{-1}$), a small subset of neutron-star LMXBs, called accreting millisecond X-ray pulsars (AMXPs), show coherent X-ray pulsations. These pulsations provide a powerful probe of the accretion process and allow measurements of the neutron-star spin period, rotational torques, and system geometry \citep{pw12}. Along with the detection of thermonuclear bursts in LMXBs \citep{sb06}, the presence of X-ray pulsations from AMXPs has shown that matter can be channeled down to the neutron star surface during periods of vigorous accretion.

In contrast to these energetic episodes, most LMXBs spend the majority of their lives in quiescence, with $L_X \lesssim 10^{34}\,\textrm{erg}\,\textrm{s}^{-1}$.  None of the known AMXPs are close enough to Earth to easily establish whether channeled accretion can occur during this state, and accretion-induced pulsations have not previously been observed in any quiescent LMXB.  The nature of the accretion flow, if any, under these circumstances is therefore poorly constrained by observations.  Generally, a strong magnetic field and rapid rotation would be expected to eject inflowing material from the system before it reaches the star, thus strongly suppressing both pulsations and luminosity \citep{ccm+98,is75}.  Although the neutron-star LMXB Cen X-4 shows evidence for accretion-induced heating of the neutron star even in quiescence, the absence of pulsations suggests that this is due to a very low neutron-star magnetic field that cannot inhibit accretion \citep{dfmp15}.

The recent transformation of the PSR~J1023+0038 system from a rotation-powered millisecond pulsar (RMSP) back to an LMXB state \citep{pah+14} provides a unique opportunity to study accretion processes at low X-ray luminosity.  The PSR~J1023+0038 system, located only 1370$\pm$40 pc from Earth \citep[parallax distance measured in radio;][]{dab+12}, consists of a neutron star in a $0.198$-day orbit with a $0.2\,M_\Sun$ main-sequence-like companion \citep{asr+09}. The observational history is complex, but PSR~J1023+0038 appeared to have an accretion disk in 2000/2001 \citep{asr+09}.  However, the X-ray luminosity of the system remained relatively low at that time ($< 10^{35}\,\textrm{erg}\,\textrm{s}^{-1}$), indicating the absence of a strong accretion-powered outburst \citep{asr+09}.  The 1.69-ms radio pulsations were discovered in 2008 \citep{asr+09}, and up until 2013 the system was observed as a RMSP, during which time all the basic system parameters were determined \citep{akh+13}.  Sometime in 2013 June, PSR~J1023+0038 changed state, disappearing as an observable RMSP \citep{sah+14} and brightening substantially in $\gamma$-rays \citep{sah+14}, optical \citep{hgs+13}, and UV \citep{pah+14}. In X-rays, the binary underwent an increase in mean X-ray luminosity from $10^{32}$ to ${\sim} 3\times 10^{33}\,\textrm{erg}\,\textrm{s}^{-1}$ (0.3--10 keV).  From then until at least 2014 November, it has had an accretion disk \citep{hgs+13} and an X-ray luminosity comparable to the known quiescent LMXBs \citep{cbc+14}.

Here we present {\it XMM-Newton} observations which show that PSR~J1023+0038 intermittently produces coherent, apparently accretion-induced X-ray pulsations during its LMXB state.  In Section~\ref{sec:observations} we present the observations and data analysis.  The results are presented in Section~\ref{sec:results}, and these are discussed in Section~\ref{sec:discussion}.

\section{Observations and analysis}
\label{sec:observations}
Motivated by this remarkable state change, we observed PSR~J1023+0038 with the \emph{X-ray Multi-Mirror-Newton} telescope. Specifically, PSR~J1023+0038 was observed with \textit{XMM-Newton} on two separate occasions: starting on 2013 November 10 for 134 kiloseconds (Observation ID~0720030101) and again on 2014 June 10 for 115 kiloseconds (0742610101). For both observations, the European Photon Imaging Camera (EPIC) pn detector was used in fast timing mode, which permits a 30 $\mu$s time resolution but at the expense of one imaging dimension, which is used for fast read-out.  Both EPIC MOS cameras were employed in small window mode to minimize the deleterious effect of photon pile-up. For all three detectors, the thin optical blocking filter was in place.

The data reduction and extraction were performed with the Science Analysis Software (SAS\footnote{The \textit{XMM-Newton} SAS is developed and maintained by the Science Operations Centre at the European Space Astronomy Centre and the Survey Science Centre at the University of Leicester.}) version {\tt xmmsas\_20130501\_1901-13.0.0}. The EPIC events were filtered using the recommended flag and pattern values.

The source events from the EPIC pn fast timing data were extracted using a region of half-width 6.5 pixels in the imaging (RAWX) direction centered on row 37. This translates to an angular size of 27$''$, which encircles $\sim$87\% of the energy from the point source at 1.5 keV. The MOS1/2 source events were obtained from circular regions of radius 36$''$, which enclose $\sim$88\% of the total point source energy at 1.5 keV. For the variability analysis and pulsation searches, the photon arrival times were translated to the solar system barycenter using the DE405 solar system ephemeris and the pulsar position with right ascension 10:23:47.687198 and declination +00:38:40.84551 calculated from the astrometric solution given in \citet{dab+12}.  

The individual binned, background-subtracted, exposure-corrected light curves for the MOS1/2 and PN were extracted using the SAS command {\tt epiclccorr}. The total EPIC light curve was then obtained by summing the three light curves during the periods when all three telescopes acquired data simultaneously. Both the 2013 November and 2014 June data exhibit multiple instances of enhanced background due to soft proton flares. Since PSR~J1023+0038 is on average relatively bright in its accreting state, the intervals of high background were not discarded. Instead the flares were removed in the background subtraction. Comparison between the final source light curve and the background light curve revealed that the background flares were no longer present  in the data. This indicates that any remaining bright flares are not due to background but are intrinsic to the source. For the fast timing photon lists, we determined time ranges corresponding to soft proton flares by thresholding a 10-s binned light curve extracted from an off-source region; photons arriving during these time ranges were not used in pulsed flux and profile calculations.

PSR~J1023+0038 has a precisely known rotational ephemeris, which predicts the pulsar's rotational phase \citep{akh+13}.  However, the pulsar exhibits non-deterministic orbital period variations \citep{akh+13}. We therefore computed photon arrival phases for a sequence of ephemerides, each constructed from the last known radio ephemeris by varying time of the ascending node ($T_{\rm asc}$) within a range of $\pm 20\,\textrm{s}$ in steps of $0.1\,\textrm{s}$, and then applied the $H$ test \citep{jrs89} to each set of phases.  
We searched in the pulsar's spin period as well, but obtained only marginal improvements in detection significance. Uncertainties in the relative timing aboard \emph{XMM-Newton} \citep{mkc+12} make it impossible to provide physically interesting constraints on the spin period using the existing data.

Further results from these X-ray observations, combined with other multiwavelength data, will be presented in \citep[][]{bab+14}. There we also describe in detail the spectral fitting and flux calculations. Briefly, we classified the photons based on luminosity mode (see section~\ref{sec:results}) and carried out spectral fitting in each mode and for each of our two epochs. We found very similar results in the two epochs, and all three modes were adequately fit with power laws (although there is some evidence for a more complicated spectrum in the high mode) with interstellar absorption ($N_H$) of $3.1(2)\times 10^{-20}\,\textrm{cm}^{-2}$. For the low, high, and flare modes in the 2013 November epoch we found photon indices of $1.82(3)$, $1.71(1)$, and $1.66(2)$, respectively; the three modes differ chiefly in their luminosities $L_X$, which are $5.4(1)\times 10^{32}$, $3.17(2)\times 10^{33}$, and $1.09(2)\times 10^{34}\,\textrm{erg}\,\textrm{s}^{-1}$ respectively. For the 2014 June epoch, the photon indices were $1.80(4)$, $1.75(1)$, and $1.74(2)$, while the luminosities were $4.6(1)\times 10^{32}$, $3.06(2)\times 10^{33}$, and $9.6(1)\times 10^{33}\,\textrm{erg}\,\textrm{s}^{-1}$ respectively. 

The above luminosities are computed for the $0.3$--$10\;\text{keV}$ energy range observed with \emph{XMM-Newton}. For theoretical purposes it would be valuable to estimate the bolometric luminosity of PSR J1023+0038, at least in its high mode. Given the power-law models we obtain from spectral fitting, one must assume spectral cutoffs to obtain a finite bolometric luminosity. Unfortunately no such cutoffs have been observed. Hard X-ray observations with NuSTAR show there is no cutoff below 79 keV, and $\gamma$-ray observations with Fermi show there must be a break of some kind below about 100 MeV \citep{tya+14}. Assuming that the power law begins in the optical and ends somewhere between these two, we compute a range of possible bolometric luminosities in the high mode, from $1.4\times 10^{34}$ to $1.1\times 10^{35}\,\textrm{erg}\,\textrm{s}^{-1}$. It may be possible to narrow this range somewhat by using the companion as a bolometer. \citet{ta05} studied the companion irradiation in the MSP state and found that the amount of heating seen then was consistent with a $2\;L_\Sun$ isotropic irradiation flux. In the accretion-disk state, the companion heating is presumably due to the emission from the disk and pulsar. The light curves in \citet{cbc+14} give us a measurement of the companion brightness (plus a constant contribution from the disk). If we treat the $V$ and $g$ filters as roughly equivalent, assume that the dark side of the companion has not brightened, and assume that the companion's total brightness is proportional to the flux in these filters, then we find that the companion's bright-side heating has increased by a factor of ${\sim}3$. This corresponds to a bolometric luminosity from the accretion disk and pulsar of ${\sim}2\times 10^{34}\;\text{erg}\;\text{s}^{-1}$, placing it towards the lower end of the above range.  Given the very considerable uncertainty in the bolometric luminosity of J1023, in Section~\ref{sec:discussion} we will compare X-ray luminosities (rather than bolometric) to other sources where possible (for example, in Figure~\ref{fig:amxps-plot}).

We also observed PSR~J1023+0038 with the Karl G. Jansky Very Large Array (VLA) a total of 13 times under the project codes 13B-439 and 13B-445, at frequencies ranging from 1 to 18 GHz.  A standard phase referencing setup (using the calibrator source J1024$-$0052) was used for all observations, and data reduction was performed with the standard VLA pipeline in the Common Astronomy Software Applications \citep[CASA;][]{mws+07} package.  PSR~J1023+0038 was detected in all epochs except one, appearing as an unresolved, variable, flat-spectrum ($-0.5 \lesssim \alpha \lesssim 0.5$, where luminosity $\propto \nu^\alpha$) source.  Along with the flat radio spectrum, the rapid variability observed (factor-of-ten change in flux density in 30 minutes) is indicative of outflowing material; similarly variable, flat-spectrum emission in other LMXB systems is attributed to a partially self-absorbed synchrotron emission from a compact jet \citep[for example][]{mf06}.  Further results from these radio observations will be presented in \citep[][]{dmm+14}. 

\section{Results}
\label{sec:results}
\begin{figure}
\centering 
\plotone{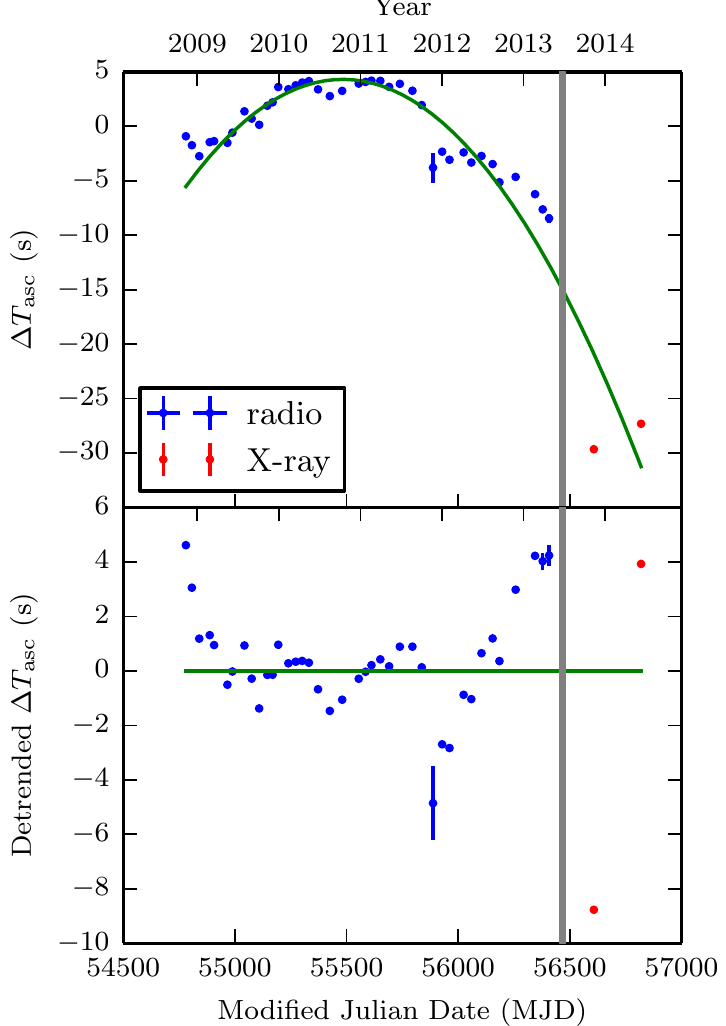}

\caption{\label{fig:T0-comparison} Variations in the time of the ascending node of the binary orbit, $T_{\rm asc}$. Blue points are measurements made by \citet{akh+13} based on radio timing; horizontal bars indicate the span of time over which the orbit was fit, and vertical bars reflect an uncertainty in the fitted values that includes unmodelled orbital deviations. The green curve is a best-fit parabola through these blue points, reflecting a long-term average binary period and binary period derivative. The gray vertical bar indicates when the radio disappearance occurred \citep{sah+14}. The red points are obtained from the two {\it XMM-Newton} observations presented here by optimizing the detection significance; their uncertainties are too small to see on this scale. The top panel was computed using a constant binary period of $0.198096315\,\textrm{day}$, obtained from long-term radio timing \citep{akh+13}. The bottom panel shows the residuals after subtraction of the green parabola, corresponding to a steady orbital period derivative of $-9.05\times 10^{-12}$. }

\end{figure}

A primary goal of our observations was to search for coherent X-ray pulsations. The pulsar's orbital period variations made it necessary to search a modest number of possible orbital phase shifts (in the form of changes of $T_{\rm asc}$) for each observation.
At the best values of $T_{\rm asc}$ for our two observing epochs, the (single-trial) false positive probability for each was less than $10^{-500}$ (${\sim} 33\sigma$), indicating a statistically unambiguous detection of pulsations in each observation. The changes in $T_{\rm asc}$ we obtained are large but not unreasonable compared to those observed while the pulsar is in its RMSP state (see Figure~\ref{fig:T0-comparison}).

\begin{figure}
\centering 
\plotone{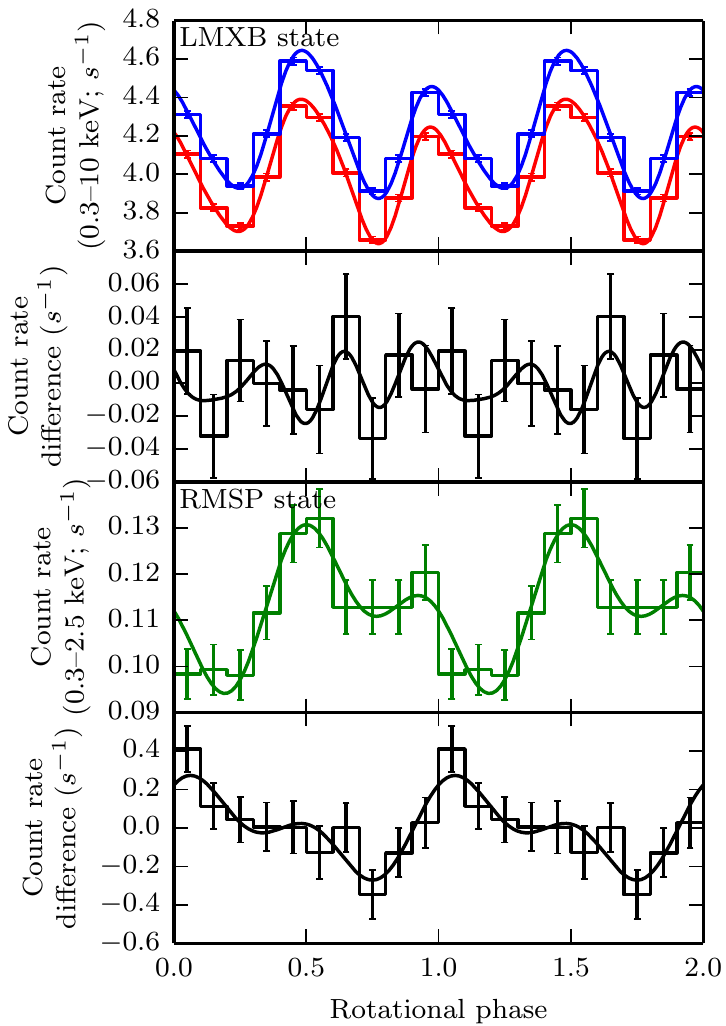}

\caption{\label{fig:profile-comparison} Top panel: the LMXB-state pulse profiles computed from the 2013 November (blue, top) and 2014 June (red, bottom) observations, when the radio pulsar was unobservable. Two rotational cycles are shown for clarity, with both a histogram and a multiple-sinusoid representation (see appendix for details). The 2014 November observation has rms amplitude $0.236(6)\,\textrm{s}^{-1}$ ($5.58(13)\%$ modulation) and the 2014 June observation has rms amplitude $0.233(6)\,\textrm{s}^{-1}$ ($5.81(14)\%$).  Second panel: the difference between the scaled LMXB-state profiles.  Third panel:  the X-ray pulse profile observed during the 2008 November observation reported in \citet{akb+10}, during which PSR~J1023+0038 was in the RMSP state (RMS amplitude $0.0099(18)\,s^{-1}$, $8.8(1.6)\%$).  Bottom panel: the difference between the LMXB-state profile (from 2014 June) and a suitably scaled version of that observed during the RMSP state (in 2008 November).}
\end{figure}

We obtained the pulse profile shapes in our two observations and found them remarkably similar (Figure~\ref{fig:profile-comparison}) but subtly different from (as well as much brighter than) the X-ray pulse profile seen in the source's previous RMSP state, where the pulsations are thought to be due to return current heating of the magnetic poles \citep{akb+10} rather than heating by channeled accretion.

\begin{figure*}
\centering 
\plotone{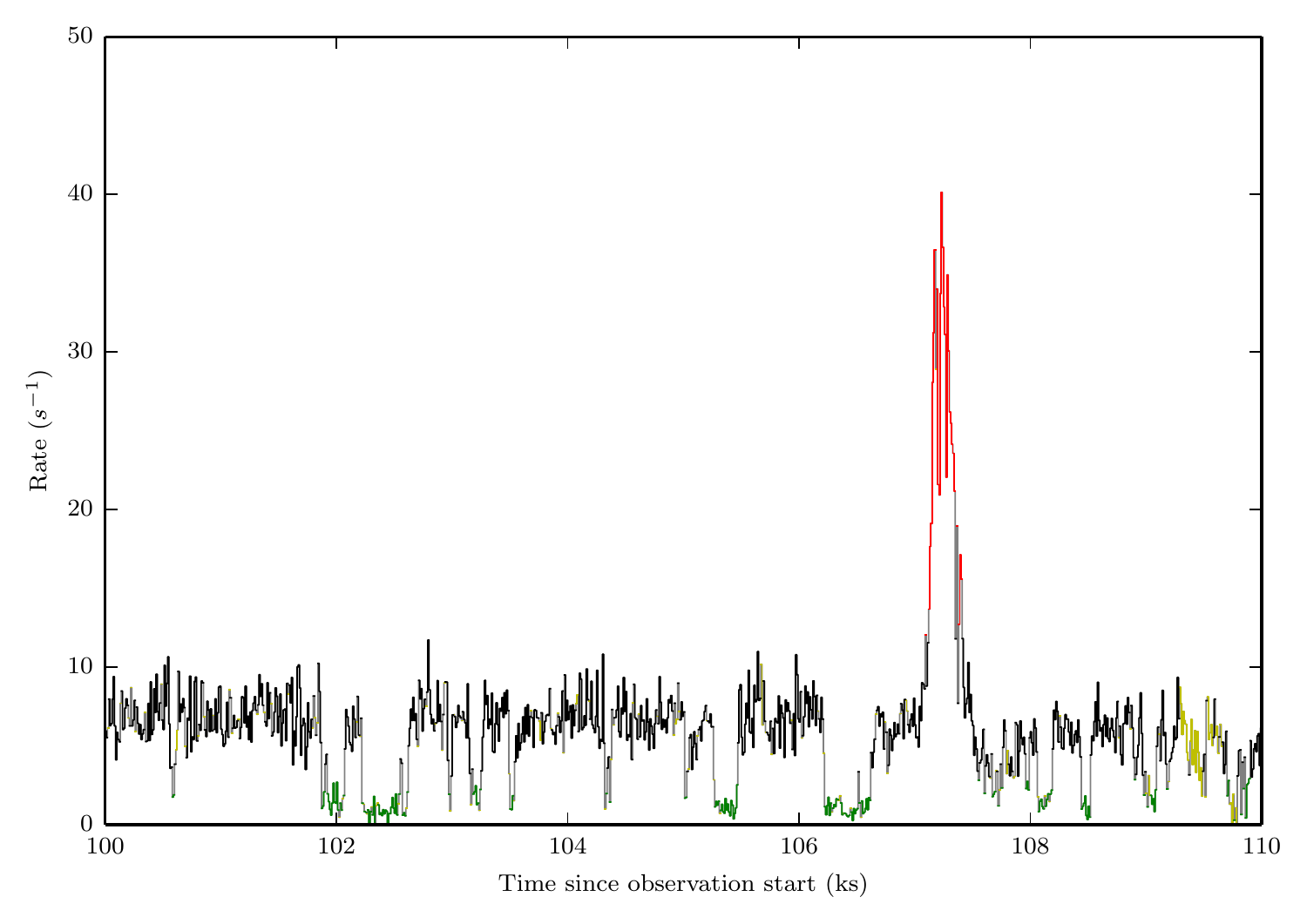}

\caption{\label{fig:lc-classify-zoom}
A short section of the combined {\it XMM-Newton} light-curve, showing the low (green), high (black), and flare (red) modes.  Times at which the mode is indeterminate are marked in grey, and times excluded from pulsation searching because of background flaring are marked in yellow. See Figure~\ref{fig:flux-distribution-lc} for how the thresholds were set, and Figure~\ref{fig:lc-classify-huge} for the whole light curve.
}
\end{figure*}

Examination of the combined \emph{XMM-Newton} light curves revealed three distinct luminosity modes\footnote{The PSR~J1023+0038 system has been observed to switch between RMSP and LMXB {\it states}.  To describe the different X-ray luminosities observed during the LMXB state, we use the term {\it mode}, that is, low, high and flare mode.} (Figures~\ref{fig:lc-classify-zoom}, \ref{fig:flux-distribution-lc} and \ref{fig:lc-classify-huge}), all of which are present during the LMXB state.  Similar behavior is seen from PSR~J1023+0038 at the higher energies observed with \textit{NuSTAR} \citep{tya+14}, as well as during the LMXB states of the two other known transitional RMSPs: PSR~J1824$-$2452I and XSS~J12270$-$4859 \citep{lbh+14,mbf+13,lina14}.  In the 0.3--10 keV photon energy range, our \emph{XMM-Newton} observations show a steady `low' mode with luminosity $\sim 5\times 10^{32}\,\textrm{erg}\,\textrm{s}^{-1}$, a steady `high' mode with luminosity $\sim 3\times 10^{33}\,\textrm{erg}\,\textrm{s}^{-1}$, and occasional, more erratic `flares' during which the luminosity typically exceeds $5\times 10^{33}\,\textrm{erg}\,\textrm{s}^{-1}$, and reaches as high as $3\times 10^{34}\,\textrm{erg}\,\textrm{s}^{-1}$.  The three modes have similar hard power-law spectra \citep[photon index ${\sim}1.7$;][]{tya+14}. The low, high, and flare modes occupy roughly 20\%, 80\%, and 1--2\% of the data, respectively, in both the 2013 November and 2014 June data sets.  Switching between the low and high modes appears to occur within $10-30\,\textrm{s}$, while the flares have a more complex structure.  The low and high mode durations are typically tens of seconds up to tens of minutes long, and the distribution of their durations follows a power law \citep{tya+14}.  No obvious periodicity or other discernible regular pattern is present.  Moreover, the luminosity in low/high mode appears to be nearly constant and equal in both the 2013 November and 2014 June observations.

\begin{figure}
\centering 
\plotone{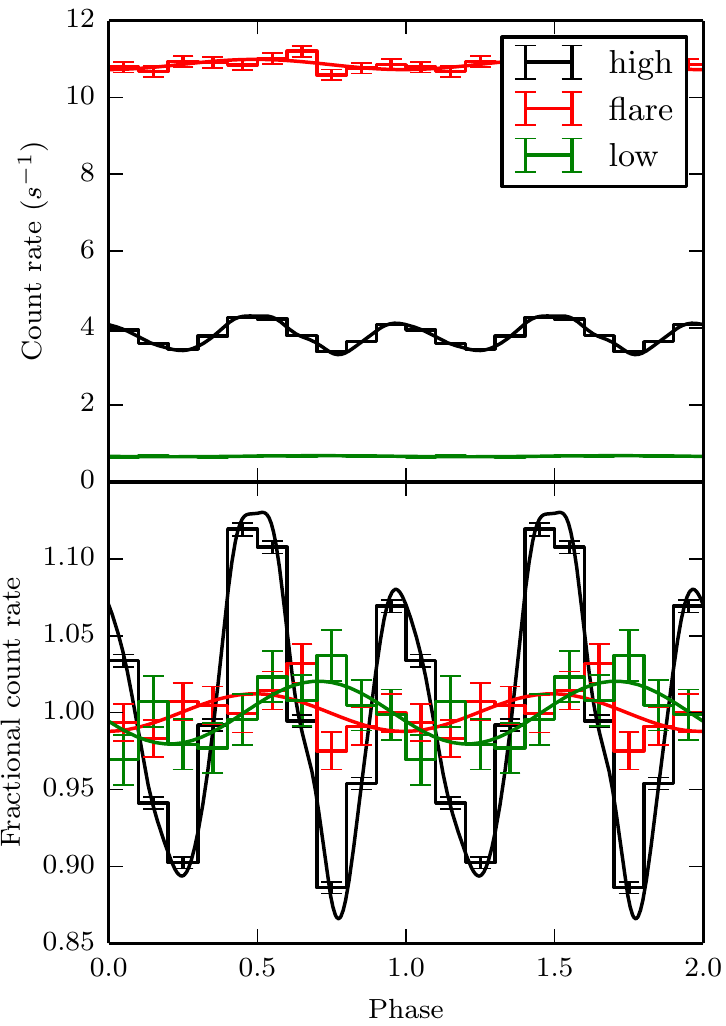} 

\caption{\label{fig:type-profile-combined} X-ray pulse profiles during the low, high, and flare modes. Each contains background-subtracted combined data from both 2013 November and 2014 June observations. The top panel shows all three normalized by count rate, while the bottom panel shows all three normalized to the same mean value. In the low mode, pulsations are not detected: the $H$ test reports a false positive probability of $0.2$; a $95\%$ confidence upper limit on the plausible pulsed fraction is about $2.4\%$ (that is, if the pulsed fraction were $2.4\%$ we would obtain a false positive probability smaller than this $95\%$ of the time). This corresponds to a maximum possible pulsed flux in the low mode of $0.016\,\textrm{s}^{-1}$. In the high mode, the pulsed flux (fraction) is $0.311(5)\,\textrm{s}^{-1}$ ($8.13(14)\%$). In the flare mode we also do not detect pulsations with a false positive probability of $0.2$, giving an upper limit on the pulsed fraction of $1.4\%$ and the pulsed flux of $0.15\,\textrm{s}^{-1}$; note that this $95\%$ upper limit on the pulsed flux is \emph{less} than the pulsed flux during the high mode.}
\end{figure}
The majority of the photons were collected during the high mode, where we find that the 0.3--10 keV X-rays are pulsed with a root-mean-square pulsed fraction of $8.13(14)\%$ (count rate $0.311(5)\,\textrm{s}^{-1}$; see also Figure~\ref{fig:type-profile-combined}). In the low mode we detect no significant pulsations, and set a $95\%$ confidence upper limit on the pulsed fraction of $2.4\%$ (count rate $0.016\,\textrm{s}^{-1}$). This implies that if any pulsations are present in the low mode, they must form a substantially smaller luminosity fraction (and much smaller absolute luminosity) in the low mode than they do in the high mode.  Pulsations are also not detected during the flares, with a $95\%$ upper limit on the pulsed fraction of $1.4\%$ (count rate $0.15\,\textrm{s}^{-1}$): if any pulsations are present during the flares, they must be weaker in absolute luminosity than during the high mode.  In other words, the flares are not simply the addition of unmodulated emission to the high mode; the pulsations appear to be suppressed during flares.

\section{Discussion}
\label{sec:discussion}
\begin{figure}
\centering 
\plotone{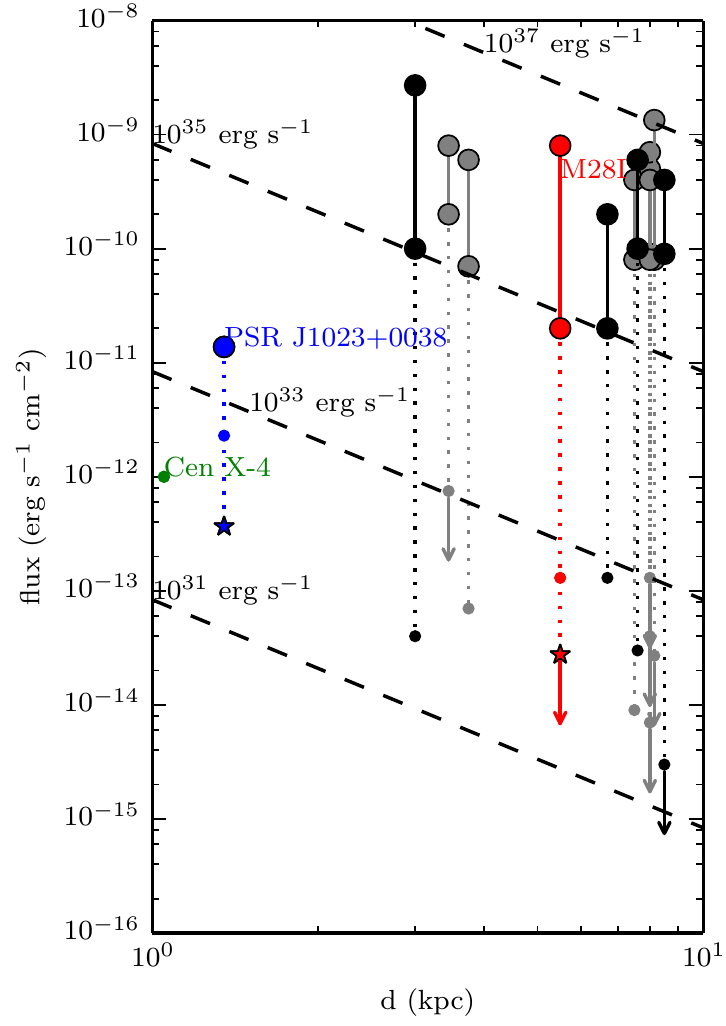}
\caption{\label{fig:amxps-plot}
Approximate X-ray fluxes of AMXPs. Large circles connected by solid lines indicate the range of fluxes at which coherent X-ray pulsations have been detected, while dotted lines and small circles indicate the range of quiescent flux levels (arrows indicate where only an upper limit on quiescent flux is available). Stars indicate X-ray fluxes during a RMSP state.  The horizontal axis is distance, generally the largest uncertainty, and the dashed diagonal lines show constant luminosities. Marked in red is the transition system PSR~J1824$-$2452I (also called M28I), and in blue is PSR~J1023+0038 (in both cases the distances are well known). For PSR~J1023+0038, the large circle shows the high-state flux (where pulsations were detected) and the small circle shows low-state flux (where no pulsations were detected).  Cen X-4 is marked in green.  AMXPs with relatively good distance estimates are marked in black, whereas those where the distance is more uncertain are marked in grey. For details and references see Table~\ref{table:amxps}.
}
\end{figure}
The detection of X-ray pulsations strongly suggests that PSR~J1023+0038 undergoes channeled accretion onto the stellar surface (at the magnetic poles) but at an X-ray luminosity of $3\times 10^{33}\,\textrm{erg}\,\textrm{s}^{-1}$ --- a level that is considered quiescent for normal LMXBs, and which is close to two orders of magnitude less than the luminosities at which AMXPs have previously been seen to pulsate ($\gtrsim 10^{35}\,\textrm{erg}\,\textrm{s}^{-1}$; see Figure~\ref{fig:amxps-plot} and Table~\ref{table:amxps}).  The observed pulse profile shapes in AMXPs are a consequence of the hot spots on the surface of the rotating star or shocks barely above it.  PSR~J1023+0038's pulsation characteristics are strikingly similar to other, higher-luminosity AMXPs: namely, a low pulsed fraction, roughly sinusoidal pulse shape, and a fairly hard spectrum \citep{pw12}.  The large duty cycle of the high-luminosity mode (${\sim}70\%$), the only one to exhibit X-ray pulsations, suggests that the channeled accretion flow in PSR~J1023+0038 is able to reach the magnetic polar caps on the stellar surface the majority of the time, but with frequent, short interruptions.

This unusual combination of phenomena appears not to be unique. While this paper was in the refereeing process, \citet{pmb+14} reported the discovery of coherent X-ray pulsations from another transitioning RMSP, XSS~J12270$-$4859, with very similar behaviour: presence in the high but not low or flare modes, system X-ray luminosity of $10^{33}$--$10^{34}\,\textrm{erg}\,\textrm{s}^{-1}$, double-peaked profile, and pulsed fraction of $7.7(5)\%$.

Theoretical models of accretion onto a magnetized neutron star must therefore account for the entirety of the rich phenomenology of PSR~J1023+0038 and XSS~J12270$-$4859: the rapid, stochastic switching between the three discrete luminosity modes --- low, high, and flare --- two of which maintain a fairly narrow range of luminosities on time scales ranging from minutes to several hours (and are also remarkably similar between observations separated by 7 months), and only one of which permits detectable X-ray pulsations.

The simplest model for accretion regulated by a stellar magnetic field assumes that a geometrically thin disk is truncated at the `magnetospheric radius', where the ram pressure of the gas balances the pressure of the magnetic field \citep{pr72}: $r_{\rm m} \simeq (B_*R_*^3)^{4/7}(GM_*/2)^{1/7}\dot{M}^{-2/7}.$ This radius is compared with the `co-rotation radius' ($r_{\rm c} \equiv GM_*/\Omega_*^2 = 24\,\textrm{km}$ for PSR~J1023+0038), the location where the orbital velocity matches the star's spin. When $r_{\rm m} > r_{\rm c}$, the star spins faster than the flow and inhibits accretion, instead expelling incoming gas from the system \citep[the `propeller' regime;][]{is75}. For PSR~J1023+0038, the transition to the propeller regime occurs at $\sim 5\times 10^{-10}\;M_\Sun\;\textrm{yr}^{-1}$ and $8\times 10^{35}\;\textrm{erg}\;\textrm{s}^{-1}$ --- substantially higher than the luminosity of PSR~J1023+0038 in its high-luminosity mode. For comparison, assuming the entire high mode X-ray flux comes from accretion onto the star's surface, we obtain a mass flow rate of $9\times 10^{-13}\;M_\Sun\;\textrm{yr}^{-1}$, almost three orders of magnitude lower and only $\sim 10^{-4}$ of the Eddington accretion rate. Even the highest plausible high-mode bolometric luminosity, $10^{35}\,\textrm{erg}\,\textrm{s}^{-1}$, corresponds to only $10^{-11}\,M_\Sun\,\textrm{yr}^{-1}$, still insufficient to permit accretion onto the surface in this model.

Our detection of pulsations at quiescent luminosities shows that the simple accretion/propeller picture is incomplete. There are several possible theoretical resolutions, each of which suggests intriguing new insights for accretion physics at low luminosities. The first is that the inner accretion flow could be `radiatively inefficient', meaning that the accretion flow changes from an optically thick disk to an optically thin, geometrically thick disk \citep{rbbp82} and most of the accretion energy goes into the kinetic energy of ejected material rather than being radiated away (so the accretion rate is much higher than the luminosity would suggest). In this case the accretion rate could be high enough to put $r_{\rm m}$ near $r_{\rm c}$, so that in the high mode some small fraction of the material can accrete onto the star. Alternately, the coupling between magnetic field lines and the highly conducting accretion flow may not be efficient. This can lead to diffusion of gas inward via Rayleigh-Taylor instabilities \citep{kr08}, or a large-scale compression of the magnetic field, both of which have been seen in numerical simulations~\citep{rukl05,ukrl06,zf13}. As a result, although the naively-computed magnetospheric radius remains well outside the corotation radius, a fraction of the gas is nevertheless able to overcome the magnetic field's centrifugal barrier and accrete onto the star. 

Finally, a propeller may not form at all. Propeller-mode accretion can only eject infalling material if the magnetic field at $r_{\rm m}$ rotates significantly faster than the disk, which occurs well beyond the co-rotation radius \citep{st93}. If this is not the case, gas stays confined in the inner part of the flow, and $r_{\rm m}$ is `trapped' near corotation~\citep{ds10,ds12,ss77}. The gas flow serves as a large reservoir of matter whose mass provides the inward pressure to balance the magnetic stress and results also in accretion episodes in spite of a low net accretion rate. The switching between low and high mode might be the result of transitions between a non-accreting pure propeller mode and an accreting trapped-disk mode.

While the high-mode luminosity provides a bound on how much matter reaches the neutron star, it is more difficult to measure the amount of material flowing into the inner regions if some of this matter is ejected. Our radio continuum observations have detected a flat-spectrum radio source, which is strong evidence for an outflow, and which may be driven either by a strong propeller~\citep{rukl05,ukrl06,zf13} or by the intrinsic accretion flow properties (as in black holes). Further modeling is required to see whether propeller-mode accretion can, with minimal radiation, eject the vast majority of the inflowing material, leaving less than 1\% to fall on the neutron-star surface. For comparison, the accreting white dwarf AE Aquarii, considered to be the clearest example of propeller-mode accretion, appears to be allowing only $0.3\%$ of the incoming material to reach the surface, but on the other hand most of the soft X-ray luminosity is produced in the inflow before ejection or accretion onto the surface \citep{om12}.  

PSR~J1023+0038 likely still has more to offer in the quest to unravel the true nature of low-level accretion onto neutron stars; in particular, continued observations yielding a measurement of the change in spin-down rate caused by this current episode will be crucial in revealing the ratio between accreted and ejected material.

\acknowledgements
The results presented were based on observations obtained with \textit{XMM-Newton}, an ESA science mission with instruments and contributions directly funded by ESA Member States and NASA. The National Radio Astronomy Observatory is a facility of the National Science Foundation operated under cooperative agreement by Associated Universities, Inc.  A.M.A. was funded for this work through an NWO Vrije Competitie grant to J.W.T.H.. A.P. acknowledges support from an NWO Vidi fellowship.  J.W.T.H. and C.B. further acknowledge funding from an NWO Vidi fellowship and ERC Starting Grant `DRAGNET' (337062; PI J.W.T.H.). A.T.D. acknowledges support from an NWO Veni Fellowship.  V.M.K. acknowledges support from an NSERC Discovery Grant and Accelerator Supplement, the FQRNT Centre de Recherche Astrophysique du Qu\'ebec, an R. Howard Webster Foundation Fellowship from the Canadian Institute for Advanced Research (CIFAR), the Canada Research Chairs Program and the Lorne Trottier Chair in Astrophysics and Cosmology.

Facilities:
\facility{XMM}
\facility{VLA}

\appendix

\begin{deluxetable*}{lcccc}
\tabletypesize{\scriptsize}
\tablecolumns{6}
\tablewidth{0pt}
\tablecaption{ AMXP flux ranges and distances \label{table:amxps}}
\tablehead{
\colhead{Name} &
\colhead{$d$ } &
\colhead{Observed pulsation flux } &
\colhead{Quiescent flux } &
\colhead{Refs.} \\
\colhead{}&
\colhead{($\textrm{kpc}$)} &
\colhead{($10^{-10}\textrm{erg}\,\textrm{s}^{-1}\,\textrm{cm}^{-2}$) } &
\colhead{($10^{-14}\textrm{erg}\,\textrm{s}^{-1}\,\textrm{cm}^{-2}$)} &
\colhead{}}
\startdata
\bf PSR J1023+0038 & \bf  $1.4$  & \bf  $0.14$ $^a$ & \bf  $229.3$ $^a$ & \bf This work \rm \\
Cen X-4 &  $0.9$--$1.2$  (?) &  \nodata &  $100.0$ $^b$ &  (1) \\
IGR J00291+5934 &  $2.5$--$5.0$  (?) &  $0.70$ -- $6.00$ $^c$ &  $7.0$ $^b$ &  (2),(3) \\
IGR J17498-2921 &  $7.6$  &  $1.00$ -- $6.00$ $^d$ &  $3.0$ $^b$ &  (4),(5) \\
IGR J17511-3507 &  $<6.9$  (?) &  $2.00$ -- $8.00$ $^e$ &  $<75.0$ $^b$ &  (6),(7) \\
PSR J1824-2452I (M28I) &  $5.5$  &  $0.20$ -- $8.00$ $^b$ &  $13.0$ $^f$ &  (8),(9) \\
NGC6440 X-2 &  $8.1$--$8.9$  &  $0.90$ -- $4.00$ $^g$ &  $<0.3$ $^b$ &  (10),(11) \\
SAX J1808.4-3658 &  $2.5$--$3.5$  &  $1.00$ -- $27.00$ $^e$ &  $4.0$ $^b$ &  (12),(3) \\
Swift J1749.4-2807 &  $5.4$--$8.0$  &  $0.20$ -- $2.00$ $^g$ &  $13.0$ $^b$ &  (13),(5) \\
Swift J1756.9-2508 &  $8.0$  (?) &  $0.80$ -- $4.00$ $^h$ &  $<13.0$ $^f$ &  (14),(15) \\
XTE J0929-314 &  $>5.0$  (?) &  $0.80$ -- $4.00$ $^d$ &  $0.9$ $^b$ &  (16),(17) \\
XTE J1751-305 &  $>6.3$  (?) &  $0.80$ -- $13.40$ $^d$ &  $<2.7$ $^b$ &  (18),(17) \\
XTE J1807-294 &  $8.0$  (?) &  $0.90$ -- $7.00$ $^d$ &  $<4.0$ $^b$ &  (19),(20) \\
XTE J1814-338 &  $8.0$  (?) &  $0.90$ -- $5.00$ $^e$ &  $<0.7$ $^b$ &  (21),(3) \\
\enddata
\tablecomments{AMXP parameters used in Figure~\ref{fig:amxps-plot}. 
Distances with a question mark indicate distance estimates highly 
dependent on assumptions. Energy ranges are indicated by 
superscript letters: \emph{a}: $0.3$ -- $10.0$, \emph{b}: $0.5$ -- $10.0$, \emph{c}: $2.5$ -- $25.0$, \emph{d}: $2.0$ -- $10.0$, \emph{e}: $2.0$ -- $25.0$, \emph{f}: $0.3$ -- $8.0$, \emph{g}: $2.0$ -- $16.0$, and \emph{h}: $2.5$ -- $16.0$ keV. 
No pulsations have been detected from Cen X-4, so it is not an 
AMXP; it is included because accretion onto the surface has
been detected. References: (1) \citet{cbd+13}, (2) \citet{hgc11}, (3) \citet{hjw+09}, (4) \citet{pbf+11}, (5) \citet{dpw12}, (6) \citet{rpb+11}, (7) \citet{hdh12}, (8) \citet{fbp+14}, (9) \citet{lbh+14}, (10) \citet{pd13}, (11) \citet{hac+10}, (12) \citet{hpc+08}, (13) \citet{acp+11}, (14) \citet{pam10}, (15) \citet{psb+07}, (16) \citet{gcmr02}, (17) \citet{whh+05}, (18) \citet{rbs+11}, (19) \citet{rsb+08}, (20) \citet{cfsi05}, (21) \citet{hp11}}
\end{deluxetable*}

\section{Pulsed fluxes and fractions.} 
The $H$ test is an effective tool for detecting the presence of pulsations, but we would often like to quantify the pulsations in order to describe the degree of modulation of the X-ray flux. The obvious approach is to choose the minimum of the light curve as an estimate of the constant background flux and then give the fraction of photons above that minimum as the pulsed fraction. Unfortunately this suffers from a potentially serious statistical bias: the process of choosing the minimum of a noisy light curve unavoidably tends to underestimate the minimum of the true light curve. The uncertainty on the pulsed fraction is also large, since it is set by the uncertainty on the minimum of the light curve. We therefore use another way to describe the pulsed flux: we report the root-mean-squared modulation. If the true light curve is $f(\phi)$, we report
\[
F_{rms} = \int (f(\phi)-\bar f)^2 d\phi,
\]
where $\bar f$ is the mean of $f(\phi)$. The relationship between this and the above definition is a scaling that depends on the pulse profile shape. 

The rms pulsed flux can be reliably estimated from the Fourier coefficients of the profile using Parseval's theorem, with correction terms to reduce the bias, as
\[
F_{rms}^2 = \sum_{j=1}^m a_j^2+b_j^2-\sigma_{a,j}^2-\sigma_{b,j}^2,
\]
where $a_j$ and $b_j$ are the $j$th Fourier coefficient, $\sigma_{a,j}$ and $\sigma_{b,j}$ are their uncertainties, and $m$ is a suitable number of harmonics. If $a_j$ and $b_j$ are normally distributed around the true values, then an easy computation shows that the expected value of $F_{rms}^2$ is the true rms pulsed flux; in other words this is an unbiased estimator. Unfortunately taking the square root introduces a small amount of bias, and in fact since $F_{rms}^2$ can turn out to be negative one must report zero in some cases, but this provides a way of computing the pulsed flux that has low uncertainty and little bias.

A priori $m$ can be chosen freely without biasing the result, but too small an $m$ will distort the estimated profile by failing to include real structure, while too large an $m$ will distort the estimated profile by including Fourier coefficients dominated by noise. There is a simple criterion for choosing $m$: select the value $m$ that maximizes $\left(\sum_{j=1}^m (a_j/\sigma_{a,j})^2+(b_j/\sigma_{b,j})^2\right) - 4m$. This criterion asymptotically selects the $m$ that minimizes the expected mean integrated squared error between the estimated profile and the true profile \citep{jrs86}. This is also the value of $m$ used in the $H$ test, and it is suitable for graphical profile representations as well as choice-free pulsed flux estimation. 

To construct the smooth profile approxmations in Figures~\ref{fig:profile-comparison} and \ref{fig:type-profile-combined}, we computed the empirical Fourier coefficients of the set of photon phases ($A_m = \sum_k e^{2\pi i m \phi_k}$), truncated them at the value of $m$ obtained from the above procedure, and plotted the resulting curve.

\begin{figure}
\centering 
\plotone{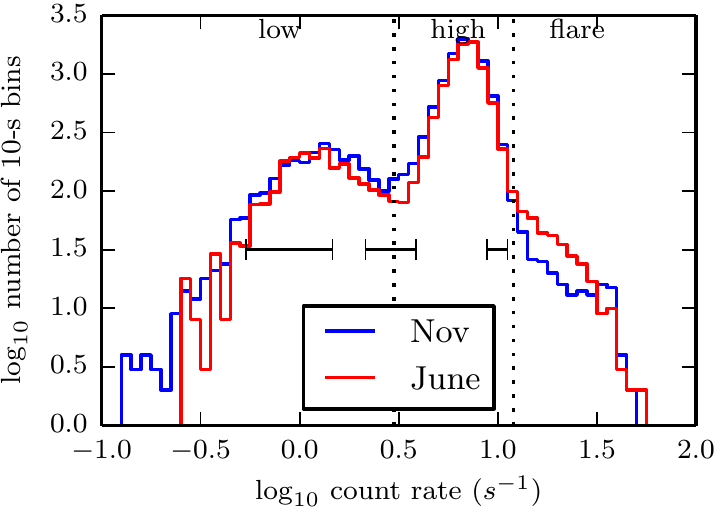} 

\caption{\label{fig:flux-distribution-lc} Distribution of fluxes in the 2013 November and 2014 June observations. All pn and MOS photons from both observations were combined to form an exposure-corrected light curve with 10-second time bins, and histograms of the resulting count rates are plotted. Horizontal error bars show the horizontal smearing due to Poisson noise in the light curve bins. Vertical dotted lines show the cuts we chose between low, high, and flare modes.}

\end{figure}

\begin{figure*}
\centering 
\plotone{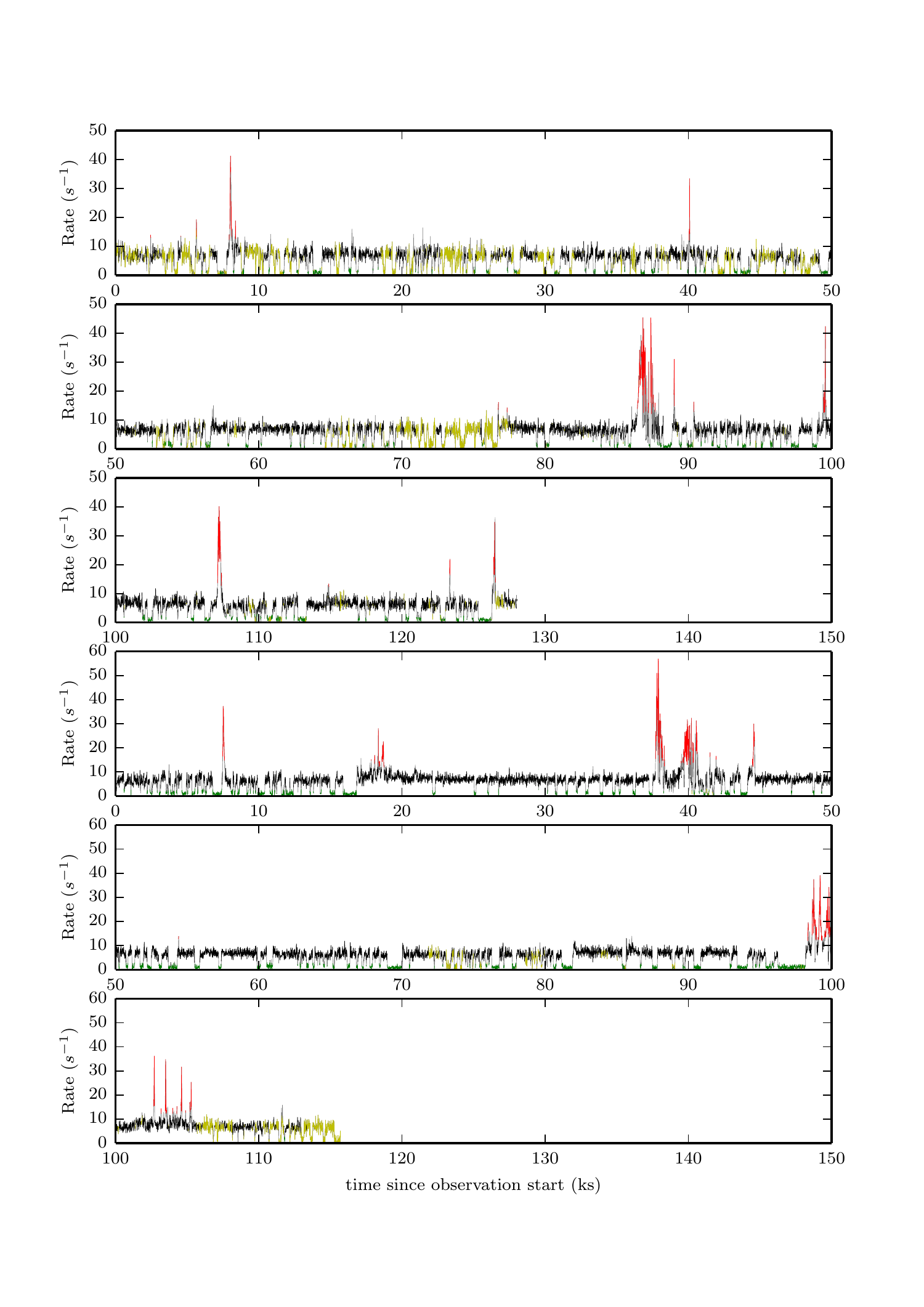} 

\caption{\label{fig:lc-classify-huge} Combined X-ray light curve showing classification of times into low, high, flare, ambiguous, or background flaring. This light curve is formed using exposure-corrected and background-subtracted data from the pn and both MOS cameras in 10-s time bins. Regions marked in green are classified as low-mode, black as high-mode, and red as flares; the gray regions were treated as ambiguous, and the yellow regions suffered from background flaring, and neither were used in pulsation searches. The top three panels of this light curve are from the 2013 November observation, while the bottom three are from the 2014 June observation.}

\end{figure*}

\section{Profile comparison.}
For the purposes of comparing the X-ray profiles between our two LMXB-state observations and to the RMSP-state X-ray profile, we formed profiles using the above optimal numbers of harmonics and aligned all three using cross-correlation. To form the difference profiles in Figure~\ref{fig:profile-comparison} we also scaled and offset one of the profiles to the least-squares best-fit values. We formed the smoothed profiles in Figure~\ref{fig:type-profile-combined} using the same Fourier-domain approach.

\bibliography{refs}

\bibliographystyle{apj}

\end{document}